# Transient Polarization and Dendrite Initiation Dynamics in Ceramic Electrolytes

Rajeev Gopal[a], Longan Wu[a], Youngju Lee[a], Jinzhao Guo[b], and Peng Bai[a,c,*]

[a] Department of Energy, Environmental and Chemical Engineering, Washington University in St. Louis, 1 Brooking Dr, St. Louis, MO 63130, USA.

[b] Department of Mechanical Engineering and Materials Science, Washington University in St. Louis, 1 Brooking Dr, St. Louis, MO 63130, USA.

[c] Institute of Materials Science and Engineering, Washington University in St. Louis, 1 Brooking Dr, St. Louis, MO 63130, USA.

* Correspondence to: pbai@wustl.edu

**Abstract**

Solid-state electrolytes, by enabling lithium metal anodes, may significantly increase the energy density of current lithium-ion batteries. However, similar to their liquid counterparts, these hard and stiff electrolytes can still be penetrated by soft Li metal, above a critical current density (CCD). The prevailing method to determine the CCD employs step-wise galvanostatic cycling, which suffers from inconsistent active interfacial areas due to void formations after repeated stripping and plating, leaving large variance in the reported data that preclude precision understandings. Here, we combine a one-way polarization technique with electrochemical impedance spectroscopy to uncover, for the first time, the existence of significant polarization dynamics in ceramic electrolytes. In contrast to the diverging transient current due to metal penetration, the current peaks we observed suggest a diffusion-limited mechanism that follows the classic Randles-Sevcik equation for analyzing the diffusion-limited processes in liquid electrolytes. Our results allow a rigorous self-consistent analysis to reveal that the CCD is a diffusion-limited current density, while the system-specific limiting current density for ceramic electrolytes is still lower than CCD, which suggests that the ion transport mechanism preceding the dendrite penetration in ceramic electrolytes is unifiable with that in liquid electrolytes.

**Broader Context**

Solid-state electrolytes have the potential to enable anode-free lithium metal batteries to achieve the highest possible energy density. In these anode-free batteries, no active anode material is needed. Instead, lithium ions in the lithiated cathode will be plated on the anode-side current collector during recharge. However, the nonuniform plating of lithium has been found to penetrate the ceramic electrolyte via elusive mechanisms. Developing precision analytical methods to understand the dynamic behaviors of lithium ions before the onset of metal penetration is critically needed in practice. Establishing the connection of the dynamics experienced in solid to that in liquid electrolytes is mutually beneficial to both systems toward developing safer batteries.

**Introduction**

Lithium-ion batteries are the key enabling technology for portable electronics and electrical mobilities. Breakthroughs for increasing the energy density are still critically needed. One promising route is the all-solid-state battery, in which a nonflammable solid-state electrolyte (SSE) is expected to stabilize both the lithium (Li) metal anode and retard the chemical degradation of intercalation cathode to achieve higher energy density, better safety, and longer cycle life. However, penetrations through ceramic SSEs, most notably the garnet-type cubic $Li_7La_3Zr_2O_{12}$ (LLZO), have been reported in various operation conditions[1–3].

One of the most commonly used criteria to characterize the SSEs is the critical current density (CCD), which is determined in step-wise galvanostatic cycling of a Li|SSE|Li symmetric cell until a sudden voltage drop toward 0 V is observed[4–6]. Multiple factors have been identified to influence the CCDs[3], including porosity[7], grain size[8], grain dopants[9], grain boundary modifications[10], sintering conditions[11,12] pellet defects[13,14], and electronic conductivity[15,16]. Common to most studies, the repeated plating and stripping generate voids at the SSE|Li interfaces, therefore causing significant contact loss even under high stacking pressures[13,17–19]. The differences between the true working current densities and the apparent current densities were not carefully quantified, leaving wide variations that impede rigorous theoretical analyses. In addition to the interfacial effects, the material and pellet properties of samples from different synthesis routes further complicated the identification of reliable correlations. Preparing ceramic materials and pellets in high consistency and large quantity and ensuring the accurate determination of true working current density, are the key challenges blocking the establishment of fundamental understandings of the CCD and the metal penetration dynamics.

Here, large quantities of highly consistent SSE samples were prepared and tested by a one-way electroanalytical technique, i.e. linear sweep voltammetry (LSV), combined with electrochemical impedance spectroscopy (EIS) measurements at different stages. We discover, for the first time, the existence, development, and recovery of transient concentration polarization in SSE pellets of various thicknesses that prove a diffusion-limited mechanism preceding the dendrite initiation.

**The consistent critical current density**

A major benefit of LSV is the absence of stripping, hence the formation of voids, at the working electrode, mitigating areal evolution commonly experienced by cycling experiments where repeated stripping and plating promotes void formation and localized current densities that can be several orders higher than the nominal apparent current density. Figure 1 shows our experimental setup and the typical results from LSV tests. To ensure high consistency of physical properties among SSE samples, ceramic pellets were fabricated using the same $Li_{6.4}La_3Zr_{1.4}Ta_{0.6}O_{12}$ (LLZTO) powder and sintered under the same conditions. Circular pellets were cut into parallel-faced pieces using a low-speed diamond saw (Fig. 1a). The miniature samples (approximately 2 mm × 2 mm × the desired thicknesses) were polished to remove any high-resistance carbonate or hydroxide surface layers and possible imperfections. Two opposite faces of the polished sample were coated with molten Li by a facile rubbing technique[20] (Supplementary Information Fig. S1). During electrochemical tests at 25 °C in an Argon-filled glovebox, a stacking pressure of 20 MPa was implemented to further ensure the intimate contact at the interfaces (Fig. 1b).

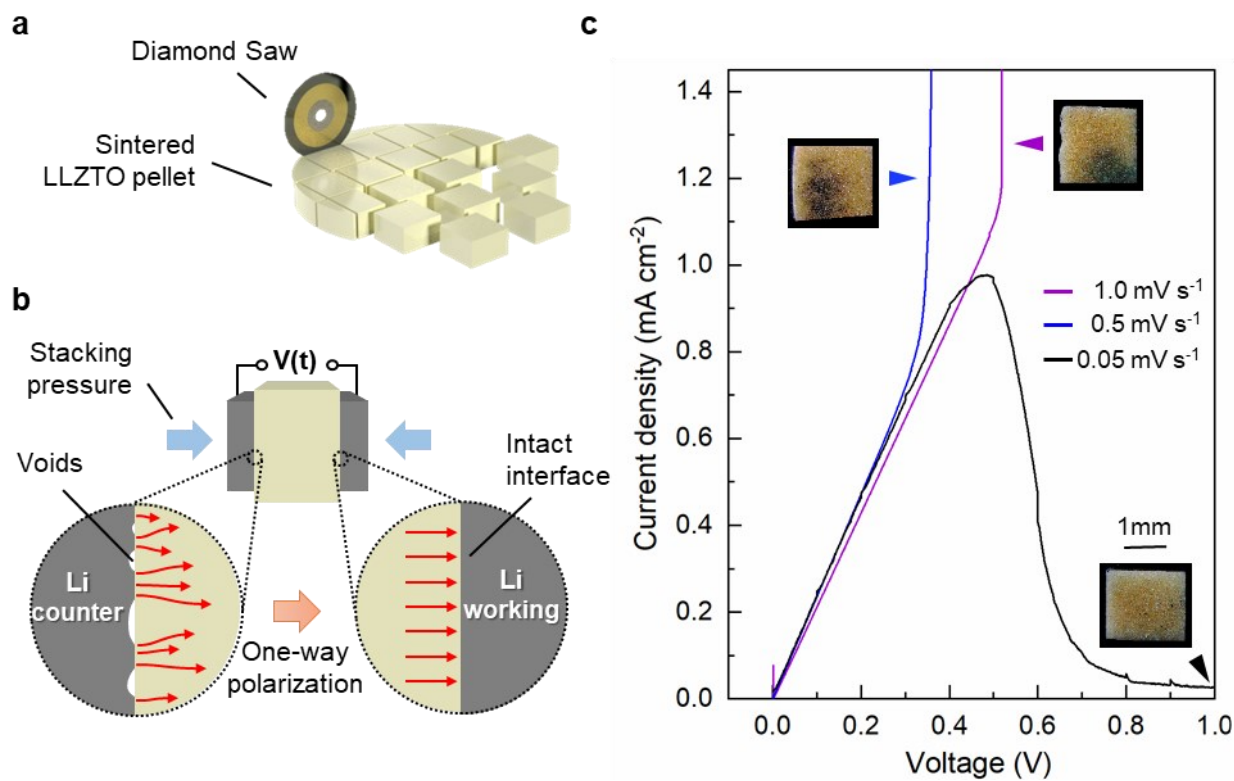

**Figure 1. Linear sweep voltammetry experiments to determine the CCD in consistent miniature samples. a,** Schematic of cutting multiple miniature samples from the same parent pellet to maximize consistency. **b,** Schematic of testing a Li|SSE|Li symmetrical cell under the one-way polarization. Red arrows in the magnified circles represent the Li-ion flux streamlines. The one-way polarization avoids the formation of voids due to repeated plating and stripping at the working electrode interface that usually occurs in the traditional method (Supplementary Information Fig. S2). Intimate interfacial contact was confirmed by a comparison of the EIS obtained via the lithium rubbing method versus gold-sputtered

electrodes, which ensured the surface is fully coated before testing. A stacking pressure of 20 MPa was maintained during the test. This working electrode interface will remain intact as stack pressure prevents potential delamination. **c,** Linear sweep voltammograms were obtained from samples with a relative density ≥ 95% at scan rates of 1, 0.5, and 0.05 mV s$^{-1}$. The inset digital photos show the surface of the SSE sample facing the working electrode, where the black regions are the residual metal penetration structures through the sample. No penetration structure was observed in samples that developed the limiting current peak, at the scan rate of 0.05 mV s$^{-1}$.

As the voltage increases, the transient current exhibits a linear Ohmic response, yielding a total conductivity consistent with impedance measurements (0.6 mS cm$^{-1}$, Supplementary Information Fig. S1). For samples tested at scan rates of 1 mV s$^{-1}$ and 0.5 mV s$^{-1}$, the transient current diverges around 1 mA cm$^{-1}$, manifesting metal penetration at the CCD (Fig. 1c and the inset photos with dark regions). As a result of much less contact loss at the working electrode, the CCDs determined here by LSV are higher than those reported in the literature studying the same material with the same configuration but using the galvanostatic cycling method. Surprisingly, however, when the scan rate was lowered to 0.05 mV s$^{-1}$, the increase of transient current was self-limited around 1 mA cm$^{-1}$ to form a peak, followed by a smooth decay toward a stable value, resembling the typical LSV responses of liquid electrolytes[21,22].

**Impedance evolution near the current peak**

The decaying current during voltage increase and the lack of penetration structure, in contrast to the diverging current at higher scan rates due to penetration (inset photos in Fig. 1c), are likely due to a diffusion-limited mechanism, but may also be attributed to the decrease of conductivity (due to concentration polarization) and the loss of interfacial contact. To decipher the dynamics behind the transient current, we repeat the LSV test but paused it intermittently every 50 mV and held at that voltage value, while a concurrent EIS test was performed to detect the bulk behavior at the paused state. Typical Nyquist plots encountered are shown in Supplementary

Information S3a and S3b. Here, a frequency range of 600 kHz to 10 Hz was adopted to ensure a short testing time of about 90 seconds, such that the state of the ceramic electrolyte will not change significantly during the voltage hold. The obtained LSV curve showed negligible deviations at these checkpoints, while the limiting current peak still emerged (Fig. 2a). A separate experiment was performed without any EIS pauses to assess any possible effects on the subsequent voltammogram. As shown in supplementary figure S12, negligible differences were observed. The choice of frequency range is sufficient to detect the targeted bulk behavior. Interestingly, the EIS spectra collected in the Ohmic region (before 250 mV) are nearly identical (Fig. 2b). Right after the current peak, however, a significant increase in impedance was discovered. Quantitative analyses by fitting the spectra to a widely used equivalent circuit model revealed the increase of impedance or decrease of conductivity, mainly occurred at the interface and the low-frequency contribution region (Supplementary Figure S3d). Here we report the small depressed semi-circle obtained at the very low frequencies as the low-frequency contribution. This region has been shown in many studies but has yet to be linked to clear physical processes. The capacitance values were found by fitting the distorted semi-circle by the constant phase element (CPE), calculated by Supplementary Information Eq. S1. Due to the high quality of our pellets, the bulk and grain boundary impedances appear as one convoluted contribution, with the resistance remaining relatively constant throughout the experiment.

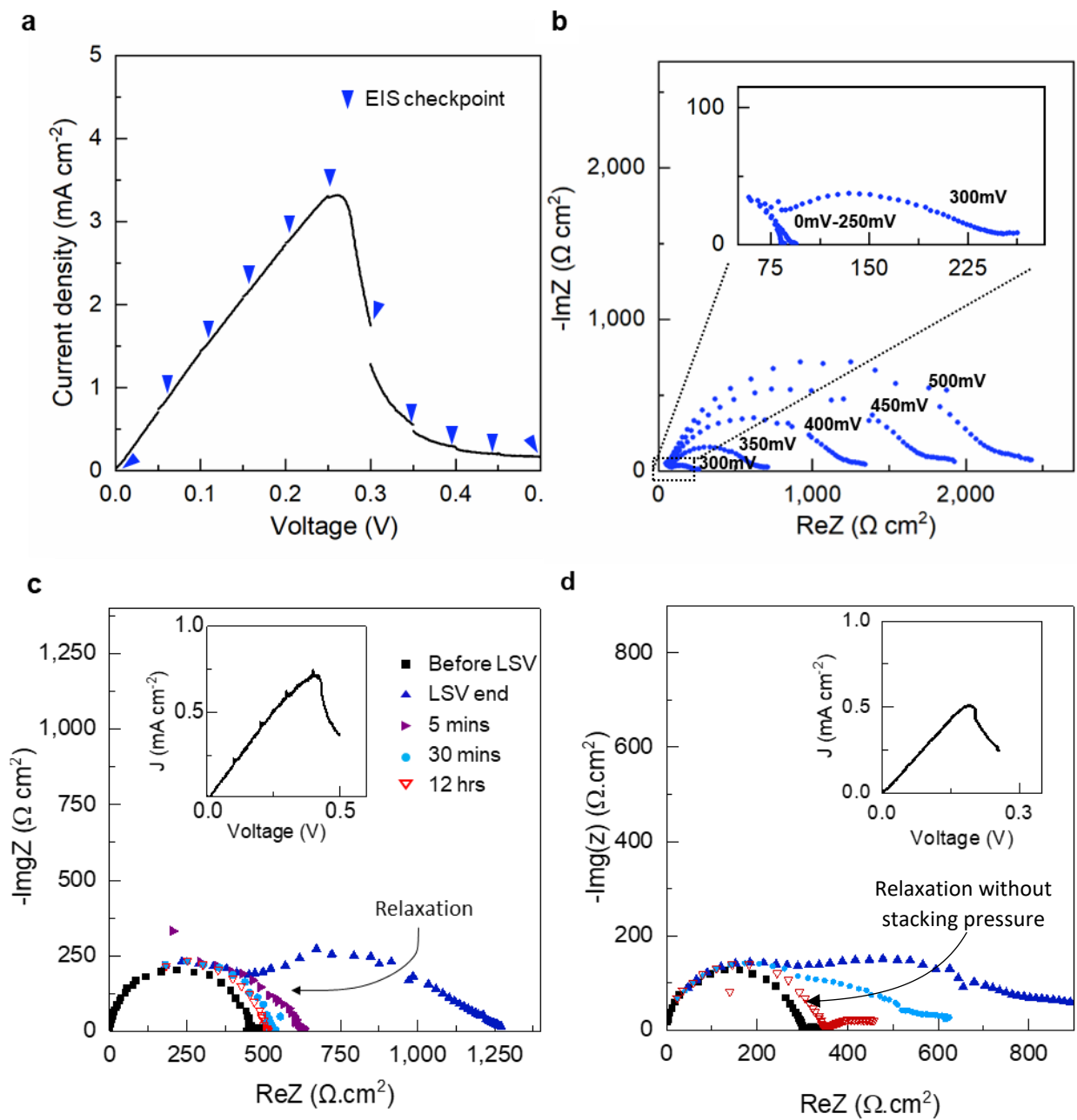

**Figure 2. Results of the combined LSV and EIS measurements of the symmetric cells**. **a,** Voltammogram showing the limiting current peak with blue arrowheads indicating the checkpoints for quick EIS measurements. **b,** Corresponding EIS spectra obtained at each checkpoint shown in panel (a). In the Ohmic region up to the limiting current peak (0 to 250 mV), changes in impedance were negligibly small. Upon further increase in voltage (> 300 mV), the low-frequency part of the spectra exhibited significant increases, which can be attributed to the grain boundary and interfacial, not the bulk, behaviors. The total resistance shown by LSV is consistent with the EIS value (Supplementary Information Table S1), displaying concordance in the two measurements. Details of the fitting are available in Supplementary Information Fig. S3. Relaxation behaviors revealed in separate LSV-EIS experiments **c,** with stack pressure (extracted data in Supplementary Information Table S2) and **d,** without stack pressure that were stopped

after the limiting peak current (additional data in Supplementary Information S4). The significant increase in impedance associated with the limiting current peak relaxed toward the original value, with a drastic decrease in the first 30 mins, followed by a much slower recovery. Insets show the corresponding voltammograms.

In liquid electrolytes, the transition from the Ohmic region to a current peak, then a stabilized current, is caused by transport limitation[22]. For SSEs, however, the common belief is that the fixed background charge from the static crystal structure ensures a near-unity transference number, such that long-range concentration polarization and therefore the transport limitation should not occur. During concentration polarization, fewer and fewer mobile lithium ions are available at the working electrode interface to maintain the original conductivity, leading to a lower subsequent total current and hence a higher measurable resistance, and eventually a diffusion limitation as reflected by the decreasing transient current. To investigate whether the impedance increase is a transient behavior that can be attributed to transport limitation, a separate LSV experiment was performed and stopped right after the current peak to monitor the impedance evolution during relaxation. These relaxation tests were performed with and without pressure to determine whether the stacking pressure may influence the interface and the subsequent relaxed impedance. For a few chosen samples, the pressure removal took less than 10 seconds and was done immediately at the end of the LSV (additional results in Supplementary Information S4). As shown in Fig. 2c and 2d, the impedance indeed recovered to the original value, and the relaxation process with or without stack pressure was the same, suggesting that the significant impedance increase was a reversible polarization and can diminish upon the removal of the electrochemical driving force. Following the basic principle of conductivity, where the concentration and mobility of the free charge carriers are the two dominant factors[23], the transient but significant changes in the impedance suggest that it was the polarization of mobile Li-ions at the interface, more specifically at the grain boundaries[5] right at the interface, that induced the lowered conductivity.

Once this polarization driving force is removed, Li-ions would return to their equilibrium concentration and the original conductivity is restored. As we shall discuss in detail in the section of *Concentration polarization preceding dendrite initiation*, the similarity of concentration polarization between the ceramic electrolytes and the liquid electrolytes are tell-tale signs of a unifiable transport mechanism leading to the Li dendrite penetration.

**Characterization of interfaces**

Li growths into, or protrusions out of, SSEs cause localized ionic flux that may lead to increased impedance and inaccurate calculation of the current densities. Before a rigorous transport analysis can be performed, interfacial contact quality between Li metal and the SSE, especially at the interface of Li plating, must be verified experimentally. Using scanning electron microscopy (SEM), cross-sectional images (Fig. 3c-f) at the working and counter electrodes from four samples, i.e. before the LSV and stopped at three different points of the LSV (Fig. 3a), were obtained. The images at 0 V (Fig. 3c) and 0.4 V (Fig. 3d) show intact interfaces fully wetted by Li metal. However, the sample stopped at 0.8 V developed clear voids at the counter electrode (Fig. 3f). For line cuts of the large two-dimensional (2D) interface. Therefore, impedance characterizations of the entire electrode, as an electrochemical response directly relevant to the operation conditions, are more reliable. The impedance spectra of the entire mm-scale electrodes are consistent with the SEM results. For the samples obtained at 0 V, 0.4 V, and 0.6 V, the significantly increased impedance can recover to the original values when the cells are fully relaxed, but the sample stopped at 0.8 V only partially recovered, leaving a wide depressed semi-circle that is consistent with the irreversible changes observed in SEM (Figs. 3b and 3f).

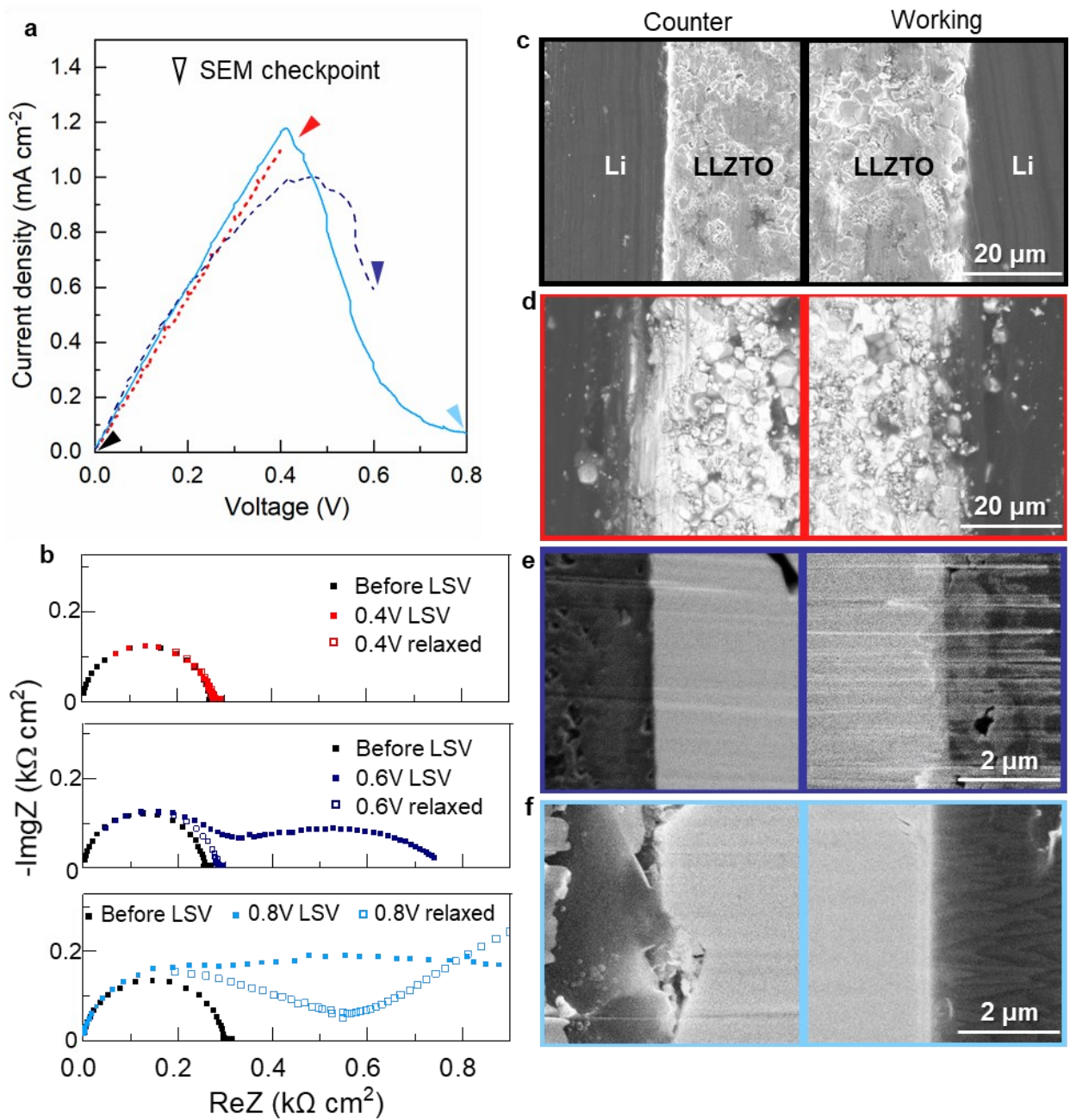

**Figure 3**. **Impedance and postmortem scanning electron microscopy (SEM) characterizations of the interfaces at different stages of the LSV. a,** Voltammograms of samples tested in separate LSV experiments that were stopped at different stages indicated by the color-coded arrowheads. **b,** Transient and relaxed impedance spectra of the three samples. For the one stopped at 0.6 V, the increased transient impedance (dark blue) can recover close to the original equilibrium state. The impedance spectra of the sample characterized at 0.8V (light blue) show much higher transient increases, which however cannot fully recover. **c-f,** Side-view SEM images of the interfaces at the counter and the working electrodes of samples

obtained at **(c)** 0 V, **(d)** 0.4 V, **(e)** 0.6 V, and **(f)** 0.8 V. The interfaces of the critical case of 0.6 V and 0.8 V were prepared by focused ion beam (FIB). No discernable voids or delamination at the local interfaces were observed for 0.6 V. Fitting results are shown in Supplementary Information Table S3. Other samples were prepared by fracturing the Li|SSE|Li cell with a sharp blade. Only in the samples obtained at 0.8 V, large gaps between the counter electrode and the SSE were observed. The results confirmed the efficacy of maintaining intact working electrode interfaces during the one-way polarization, as depicted in Figure 1.

**Effects of the bulk microstructure**

To ensure the generality and reliability of our discovery, we performed the same experiments with samples of varied relative densities. While we propose here that the transport limitation reached at the interface, more specifically at the grain boundaries at the interface, will initiate dendritic growth, this mechanism will only be uncovered if other factors do not over-shadow this mechanism. One such factor is the porosity which is tied to the relative density of the sintered pellet. A low relative density and the higher propensity of interconnected pores have been shown to promote dendrite formation and ultimately lower CCDs[4]. While the relative density appears as merely an averaged geometric property, it provides an effective single metric that quantifies the overall quality of the pellet, yet physically reflects multiple aspects of the microstructure that influence the metal penetration behavior at the CCD.

As displayed in Fig. 4a-c, an increase in the relative density led to the reduction in sizes and numbers of pits on the polished top surfaces. Samples at 99% relative density (99 rd%, versus the theoretical density of 5.491 g cm$^3$) and at 91 rd% show contrasting top-view surface features, while samples at 95 rd% appear similar to those at 99 rd%. Fractured cross-sections reveal higher internal packing density and better grain connectivity as the relative density increases. Samples at 95 rd% and 99 rd% exhibit sharp edges of intra-particle fracture, indicating well-formed grain boundaries. Higher relative densities enable higher total ionic conductivities, as reflected by the slope of the Ohmic region (Fig. 4d).

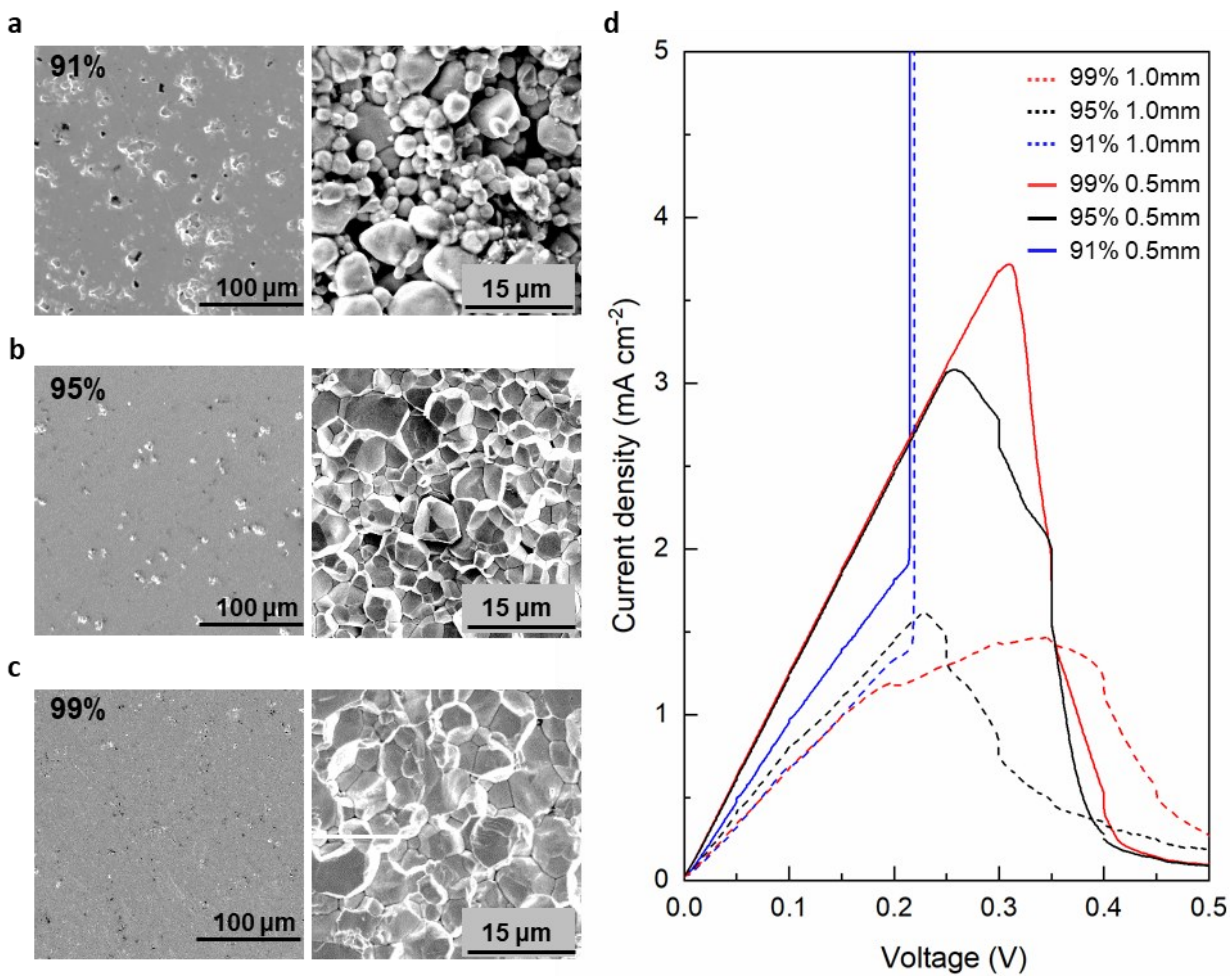

**Figure 4. Effects of structural characteristics on the transient behaviors during LSV. a-c,** SEM images of the polished top surfaces and fractured cross-sections of samples of selected relative densities. As the relative density increases, the numbers and sizes of blemishes and pores on the top surface decrease, and the cross-sections display more uniform grain sizes and tighter grain packing. **d,** Voltammograms at 0.05 mV s$^{-1}$ of samples with different thicknesses and varied relative density. Penetrative current spikes were always observed in samples with low relative densities, regardless of the thickness. In our study, the limiting current peak during LSV emerges in samples with a relative density ≥ 95%. Penetrative current spikes were not seen in samples with a relative density of 99%.

Although a relative density of 91% only indicates a slight porosity, penetrations with diverging current spikes always occur, yet at relatively low current densities. Samples at 99 rd% did not encounter any current spikes, regardless of thickness, unless stronger dynamic excitations (e.g. higher scan rates) are implemented. For samples at 95 rd%, about 10% encountered penetrative current spikes, while the rest developed the limiting current peaks. However, the CCD values determined at the onsets of the current spikes (seen in LSV with higher scan rates) are very close to those of the current peaks (seen in LSV at a scan rate of 0.05 mV s$^{-1}$). The metal penetration through samples with lower relative densities may be attributed to the excess free electrons on pore walls that can reduce mobile lithium ions under the localized electric field and thereafter promote dendritic metal penetration[15,16]. Lower relative densities may lead to percolated pores for metal filaments to grow through[4].

While the 0.5 mm-thick samples would yield an Ohmic slope steeper than the 1 mm-thick samples, as dictated by their respective total conductance, it is difficult to explain why thinner samples would also enable a much higher peak current if the key physics of the CCD is governed solely by an interfacial process. The inverse relationship between the electrolyte thickness and the peak current is true for both the 95 rd% and 99 rd% samples. Even at 91 rd%, 0.5 mm-thick samples have higher CCDs than the 1 mm-thick ones. Therefore, the total conductance, i.e. the long-range transport process, appears to play a critical role in the CCD.

**Diffusion limitations at different scan rates**

According to the Randles-Sevcik equation (Supplementary Information Eq. S2) that explains the transient current from LSV tests of liquid electrolytes[47], the peak current is scan-rate-dependent, and the scaling between the peak current ($I_p$) and the scan rate ($v$) should yield a power-law exponent of 0.5, i.e. $I_p \propto \sqrt{v}$. Here, we chose 1-mm-thick miniature pellets to repeat the LSV experiments, but at much lower scan rates. As shown in Supplementary Information Fig. S5, peak currents were still observed. The fitting of these peak current densities ($J_p$) against the scan rate ($v$) follows a power-law scaling well (Fig. 5a). Replotting the data to enforce the power-law fitting of the Randles-Sevcik equation yields a goodness of fit of $R^2=0.964$ (Fig. 5b). Our results unambiguously prove the existence of the diffusion limitation at various LSV scan rates, as revealed by both the individual peak currents and their scan-rate dependence. The quantitative fitting even suggests that the diffusion-limited mechanisms are close to that in the liquid electrolytes, based on which the Randles-Sevcik equation was developed.

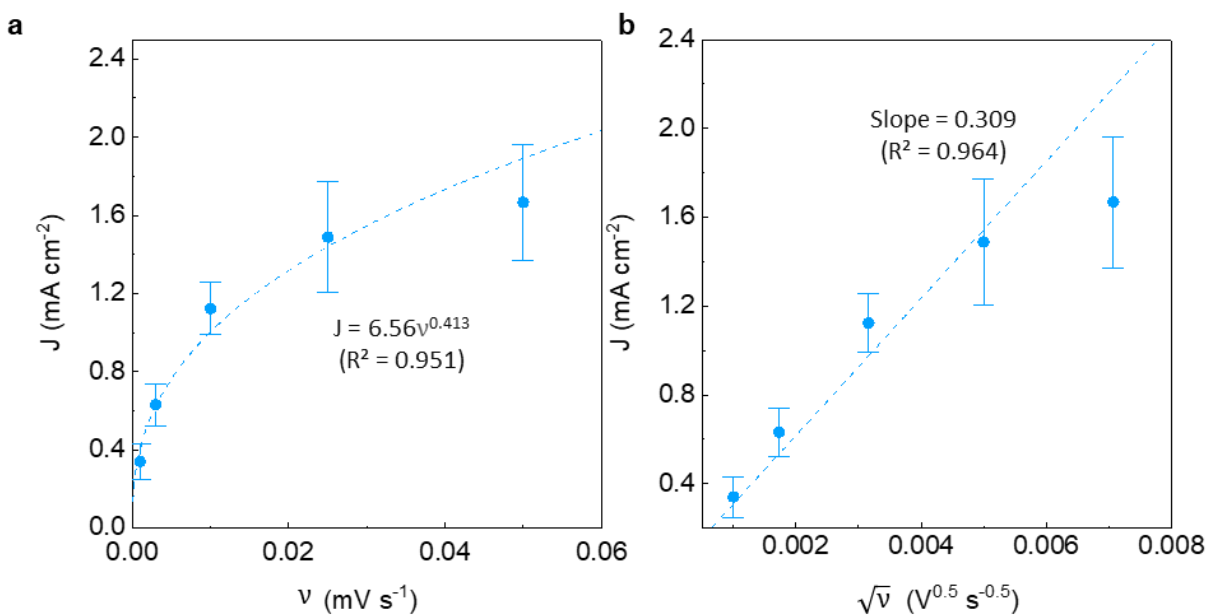

**Figure 5. Scan-rate dependence of the peak currents determined by LSV. a,** Plot of peak current densities versus scan rate for 1 mm samples shows an emergent power-law relationship, **b,** Randles-Sevcik plot to reveal the linear correlation between the peak current and the square root of the scan rate, indicating diffusion limitation. The error bars show standard deviation of the data.

Theoretically, the transient currents in response to different scan rates should eventually converge to the same limiting current at higher voltages. However, due to the interfacial damage soon after the peak (as revealed in Fig. 3f), the decaying current after the peak from each experiment is difficult to be converted to an accurate current density for cross-system comparison. As an alternative, a small enough scan rate may induce a peak current that is sufficiently close to the true limiting current. Here, we chose a scan rate of 0.003 mV s$^{-1}$ to examine the transient current and defined the obtained peak current density as the quasi-limiting current density to approximate the true limiting current density.

**Concentration polarization preceding dendrite initiation**

The limiting current density is system-specific. As demonstrated in liquid electrolytes [24–26], only when an over-limiting current density (i.e. higher than the limiting current density) is applied to the liquid systems, and the Sand's time is reached upon the complete depletion of ions at the electrode surface, would the tip-growing, fast-advancing, fractal dendrite start to form[24]. Testing the thickness-dependence of the system-specific limiting current density is another compelling approach to verify the perception that ceramic electrolytes are distinctly different from polymer[27] and liquid electrolytes[15]. Here, pellets with relative densities higher than 95% were cut and polished to reach different thicknesses of around 0.5 mm, 1 mm, 2 mm, and 3 mm. LSV experiments at scan rates of 0.05 mV s$^{-1}$ and 0.003 mV s$^{-1}$ were then performed until either a current spike indicating penetration or a current peak indicating transient limitation was observed. The scan rate of 0.05 mV s$^{-1}$ is the highest scan rate to induce the current peaks that are comparable with the CCD values, while 0.003 mV s$^{-1}$ is the lowest feasible scan rate that we chose to induce the quasi-limiting current density to probe the system properties. These two types of characteristic current density values were then plotted against the thicknesses of the samples (Fig. 6a and 6b). Corresponding voltammograms can be found in Supplementary Information Fig. S6 and S7. Note that chronopotentiometry experiments with values higher than the peak current density obtained at 0.05 mV s$^{-1}$ always led to failures due to metal penetration (Supplementary Information Fig.

S8), which further proves that the current peak reflects the same physical property as the traditional CCD does. Both the peak current density obtained at 0.05 mV s$^{-1}$ and the CCD of the traditional definition (with observed penetration) are referred to as the generic CCD.

As shown in Figs. 6a and 6b, the two types of characteristic current densities, i.e. CCD and the quasi-limiting current density, are inversely proportional to the thickness of the SSE pellets. This trend coincides with that of liquid systems, where the system-specific limiting current density[24] is strictly proportional to the reciprocal of the electrolyte thickness via $J_{lim} = 2zFDC_0[(1-t_+)L]^{-1}$. Here, $z$ is the charge number, $F$ the Faraday constant, $D$ the ambipolar diffusion coefficient, $C_0$ the bulk ion concentration, $t_+$ the lithium-ion transference number, and $L$ the thickness of electrolyte.

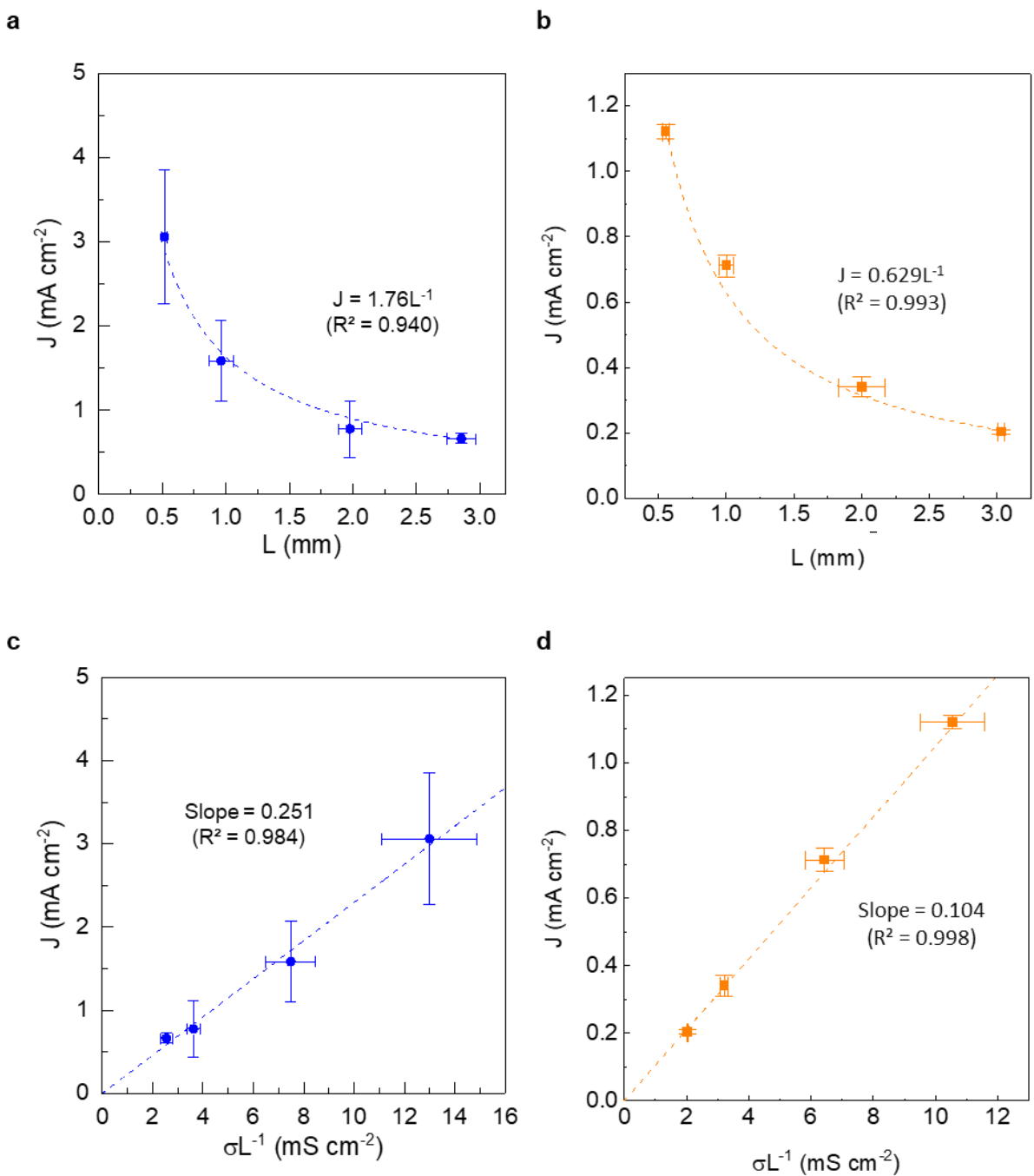

**Figure 6. Thickness-dependence of the diffusion-limited current densities. a,** Plot of CCDs against pellet thicknesses. **b,** Plot of quasi-limiting current densities against pellet thicknesses. Error bars show the standard deviation of the data set with the mean values determined by grouping similar thicknesses. Plots of (**c**) CCDs and (**d**) quasi-limiting current densities versus the thickness-

adjusted conductivities, according to Eq. (1). Full voltammograms are displayed in Supplementary Information Fig. S6 and S7.

To understand the physics behind this correlation that resembles the correlation in liquid electrolytes, we implemented the Nernst-Einstein relation to convert the total conductivity to an effective "total" diffusion coefficient, $D^\sigma = \sigma RT C_0^{-1} z^{-2} F^{-2}$. Here, $T$ is the temperature, $R$ the gas constant, and $\sigma$ is the total ionic conductivity of the LLZTO sample. Substituting $D^\sigma$ into the formula of limiting current density for dilute binary electrolyte yields a new equation:

$$J_{lim}^\sigma = \frac{2RT\sigma}{z(1-t_+)FL} \quad (1)$$

This new limiting current density based on ionic conductivity suggests a new data analysis method by plotting the characteristic current densities versus $\sigma L^{-1}$. As can be seen in both Fig. 6c and Fig. 6d, each thickness-dependent data set collapses onto a single linear fitting line. It must be pointed out that, while the CCD obtained at a scan rate of 0.05 mV s$^{-1}$ also proves a diffusion-limited mechanism, it is the one obtained at the sufficiently low scan rate that can approximate the true limiting current density of the SSE. Therefore, only the fitting slope in Fig. 6d can be used to estimate the transference number of lithium ions in our LLZTO pellets, as discussed below.

**Transference number and the mobile charge carriers**

The transference number of lithium ions in ceramic electrolytes is expected to be a value very close to 1, as lithium ions are believed to be the only significant mobile charge carrier in ceramic electrolytes. While there indeed exist works investigating the Li vacancies[28] (effectively the negative charge carriers) in ceramic electrolytes that greatly facilitate the transport of Li ions, they were not taken into account in the discussion of the transference number. The widely used definition of $t_+$ for ceramic electrolytes considers instead the leakage electrons as the only negative charge carrier. Therefore, $t_+$ is calculated as the ionic current divided by the sum of the ionic current and the electronic current. When the leakage electronic current is negligibly small, the definition leads to a near-unity $t_+$. The concentration and mobility of the Li vacancies may be low and hard to be determined[28], but their existence is critical for fast Li-ion conduction. Even if electrons must be considered in the case of ceramic electrolytes[15,16], they should not be the only negative charge carrier, excluding the contribution from the effectively negative Li vacancies[28]. The good agreement between the experimental data and Eq (1) that originated from liquid electrolytes appears to justify the binary characteristic in ceramic electrolytes.

The rigorous and proven definition of Li-ion transference number for binary liquid or polymer electrolytes[24,29–31] reminds us that $t_+$ is bounded with the ambipolar/total diffusion coefficient[32] via $D/(1 - t_+) = 2D_+$. Wherever the self/tracer diffusion coefficient of Li ions ($D_+$) can be determined, there is no need to use this secondary quantity of $t_+$ for the transport analysis. As exemplified in our Supplementary Information Table S4 and Eq. S2-S5, using the tracer diffusion coefficients reported in the literature, we were able to determine the concentration of mobile Li-ions in LLZTO to be around 20 M, by using the Randles-Sevcik slope presented in Fig. 5b, which is consistent with recent estimations[14,23]. With the concentration and the tracer diffusion coefficient, the "true" limiting current density for our LLZTO pellet is determined to be around 0.27 mA cm$^{-2}$, which is consistent with our expectation that our quasi-limiting current density obtained at the lowest feasible scan rate is slightly higher than the true limiting current density. Note that we have applied in Eq (1) a simple assumption of $DH_R^{-1} = D^\sigma$, following the suggestion[33] that the Haven ratio ($H_R$) between the tracer diffusion coefficient ($D_+$) and the one derived from the ionic conductivity ($D^\sigma$) is a constant smaller but close to 1. However, for Li-ion conducting ceramic electrolytes, the Haven ratio may vary in the range[34–36] of 0.24 – 0.7, resulting in a transference number in the range of 0.881 – 0.654, which are still more than two times higher than those determined from the liquid electrolytes.

**Discussion**

The thickness dependence revealed by Eq (1) suggests that thinner SSEs have higher intrinsic limiting current densities ($J_{lim}^{\sigma}$). As demonstrated by Hitz et al.[37], when the net flux of Li-ions (collected from the porous layer of their trilayer LLZO electrolyte) travels through a dense 15 μm layer, their cells can continuously cycle at a high current density of 10 mA cm$^{-2}$. Assuming the same correlation presented in Fig. 6b, a thickness of 15 μm yields a quasi-limiting current density of order 40 mA cm$^{-2}$. Therefore, a current density of 25% $J_{lim}^{\sigma}$ will not trigger diffusion limitation and thereafter the dendrite penetration, as long as the intimate contact at the interfaces can always be maintained. As typical separators wetted by liquid electrolytes are designed to be around 20 μm thick to avoid diffusion limitation, thinner SSEs, besides the benefits of lower total resistance and less inactive mass, will help avoid the significant concentration polarization and dendrite initiation at diffusion limitation.

Reducing grain size and therefore maximizing the grain boundary area has been demonstrated to increase the CCD. Although different fabrication methods may lead to opposite trends, our experiments show that pellets with larger grains induced earlier penetration at lower current densities (Supplementary Information Fig. S9). Our discovery that concentration polarization mainly occurs through the interfacial grain boundaries raises a question about the *true current density* at the interface and through the SSE. While using the entire geometric contact area is convenient, it is physically more appropriate to use the areal contribution from the grain boundary at the SSE|Li interface to calculate the true current density. The true ion-conducting interfacial area can be estimated by using the total length of the grain boundaries multiplied by the thickness of the space charge layer (SCL) on either side of the sharp grain boundary[38]. The true current density based only on the ion-conducting area could be at least one order of magnitude higher than the values calculated using the total interfacial area (Supplementary Information Fig. S10).

The SCL at the grain boundary may only be nanometer-thick as characterized by the Debye length[38]. Similar thicknesses for the SCL at the SSE|Li interface have also been predicted through thermodynamic analysis[39]. However, under *dynamic* electrochemical conditions, the thickness of SCL at the SSE|Li interface can extend into the SSE for hundreds of nanometers[40], as revealed by *operando* electron holography[41]. Regardless of the specific thickness, the voltage change across a length scale in the range of 1nm to 1μm would generate a significant electric field, e.g. 1.3V across 10 nm from ref[42] yields 130 MV m$^{-1}$, which can easily break down most insulating materials[43]. In liquid electrolytes, the significantly lowered salt concentration at diffusion limitation leads to decreased electrical conductivity and amplified local electric field[44], which drives fast electrokinetic flows[45] and electrode instabilities[46]. In solid electrolytes, the SCLs at the macroscopic SSE|Li interface are connected with the grain boundaries where rich defects usually facilitate fast ion transport, therefore, forming a facile ion conduction path, where apparent limiting current can occur[30] even with a unity transference number. The enormous local electric field at the diffusion limitation will lead to easy electron injection[43] through the vulnerable ion conduction path to reduce the mobile Li-ions, either at the electrode surface or at the grain boundaries in the bulk[15,16]. The critical process may also be understood effectively as the result of an induced mechanical pressure at the interface[30] that can create cracks through the electrically insulating ceramic single crystals[47,48]. While the observable dendrites or filaments grown after the diffusion limitation (induced by either high currents or large voltage bias) are indeed fundamentally different in different electrolytes[15,27], their appearances do not reject the rationale that the concentration polarization processes preceding the diffusion limitation are essentially the same for ceramic, polymer, and liquid electrolytes.

Void formations and the depletion of the intimate contact layer at the counter electrode can lead to the decrease of transient current, which may be misinterpreted as a result of transport limitation. In our experiments, however, such possibilities can be ruled out by analyzing the capacity at the current peak. Given the same Li|LLZTO interface preparation method, the same stacking pressure, and the similar thicknesses of the resulted Li metal layers, the intimate contact layer of Li metal at the counter electrode should be of similar thickness. Depletion of the similar contact layers should result in a similar characteristic capacity. However, as shown in Supplementary Figure S13, we did not observe such a common capacity at the current peaks of our LSV experiments. On the contrary, the capacity at the current peaks shows a clear dependence on the thickness of the pellets, which suggested effects from the long-range transport.

**Conclusion**

By combining a one-way polarization technique with impedance diagnosis, we have demonstrated a new method of determining the CCD that can (i) avoid the contact loss encountered in the conventional galvanostatic method, and (ii) decouple the transient behaviors to contributions from the bulk, through the grain boundary, and at the interfaces of the ceramic electrolyte. Our electroanalytical tests of miniature samples with high consistency revealed, for the first time, the existence of diffusion-limited current peaks that did not induce metal penetrations, yet manifest a scan-rate-dependent diffusion-limited behavior as predicted by the Randles-Sevcik equation for analyzing liquid electrolytes. The transient impedance spectra confirmed the development and the subsequent reversible recovery of significant concentration polarization at the macroscopic Li-electrolyte interface via the grain boundaries, but not in the bulk, of the SSE pellets. Contradictory to the prevailing belief that ceramic electrolytes are distinctly different from polymer and liquid electrolytes, our results suggest that the transport mechanism in ceramic electrolytes can be understood by the classic theory for binary liquid electrolytes, and the CCD obtained in ceramic electrolytes is essentially a diffusion-limited current density, only that the system-specific limiting current density is even lower. While thinner SSEs are preferred for the reasons of lower total resistance and higher gravimetric and volumetric energy densities, our results offer the most critical consideration as thinner SSE is vital to ensure a higher intrinsic limiting current density and CCD, such that normal operation current densities will not trigger diffusion limitation and the following lithium dendrite penetration. It is noteworthy that highly densified electrolyte pellets (relative density > 95%) are critical to prevent metal growth through percolated pores, especially in thinner electrolytes. Rational engineering of grain boundaries and promoting uniform flux distribution at the macroscopic Li/SSE interface are beneficial to avoid localized diffusion-limited metal penetrations.

**Experimental Section**

*Powder and SSE sintering:* Cubic Ta-doped LLZO (LLZTO) with the required composition of $Li_{6.4}La_3Zr_{1.4}Ta_{0.6}O_{12}$ was prepared using a solid-state technique where precursor powders of $Li_2CO_3$ (Alfa Aesar 99%), $La(OH)_3$ (Alfa Aesar 99%), $ZrO_2$ (Alfa Aesar), and $Ta_2O_5$ (Acros Organic) were dispersed in isopropanol and ball-milled at 500 rpm for 6 hrs using a zirconia milling set. These powders were added in stoichiometric amounts, except for $Li_2CO_3$, which had a 20% weight excess to account for Li loss during calcination. The resulting mixed powder was then dried overnight in a convection oven at 75 °C and hand-pressed into pellets for calcination. The pellets were then calcined in a covered MgO crucible at 1000 °C for 2 hrs. These pellets were then hand-ground with a mortar and pestle and sieved using a 32 μm mesh size. The resulting powder was collected and immediately placed in an inert argon atmosphere to prevent surface reaction with humid air. LLZTO reference powder was also purchased from Ampcera through MSE supplies and MTI, and used for tests.

To fabricate large numbers of pellets up to 96 rd%, conventional tube furnace sintering was used. The LLZTO powder was mixed with a 2% polyvinyl butyral binder purchased from Fischer Scientific in ethanol. The dried powder was uniaxially pressed at 300 MPa in a 12.7 mm diameter stainless-steel die. This green body was then isostatically pressed at 350 MPa for 10 mins. For sintering, the green body was suspended on a platinum support[49] and placed in a magnesia (MgO) crucible covered with an accompanying lid. 0.5 g of loose LLZTO powder was placed in the crucible as bed powder and sintered in a tube furnace at 1250 °C for 20 mins. A ramp rate of 3 °C min$^{-1}$ was used for the heating and cooling process. The final pellets had a tan coloration.

To achieve 99 rd% pellets, rapid induction hot pressing was used. The powder was loaded into a 12 mm graphite die and pressed at 60 MPa while simultaneously being heated to 1200 °C using an induction coil. Heating was done under an inert atmosphere of flowing argon gas. A typical ramp rate of 150 °C min$^{-1}$ was used for the heating process, with a 1 hr hold time at 1200 °C and a 10 °C min$^{-1}$ cooldown rate. The sintered pellet was extracted from the graphite die and placed in a muffle furnace at 1000 °C for 1 hr to burn off residual graphite. The final pellets had a glassy brown coloration.

The sintered pellets were then cut using a low-speed diamond saw into multiple pieces. These pieces were polished using 1200 grit and 2500 grit sandpaper followed by fine polishing on a polishing pad with a 50 nm alumina glycol solution. The samples were rinsed with isopropyl alcohol and immediately placed in an argon-filled glove box to minimize the surface reaction with air[50].

*SEM Imaging:* Thermofisher Quattro S environmental SEM was used at 10 kV accelerating voltage and 10 mm working distance. The polished top surfaces of the LLZTO and Li|SSE interface were transferred to the SEM in an airtight homemade SEM holder to minimize exposure to air and subsequent oxidation. The surface and interfacial voids were analyzed using ImageJ.

*XRD:* A Bruker d8 Advance XRD was used to analyze the crystalline phases in the LLZTO powder before sintering (Supplementary Information Fig. S11). Data was collected in increments of $0.02°$ at $15°.min^{-1}$ from 10 to $70°$ $2\theta$ at 40 kV and 40 mA using Cu $K_\alpha$ radiation. PDF 04-018-9024 was used as the reference and analyzed using Bruker Diffrac.

*Electrochemical Measurements*: A Gamry Reference 600+ was used for the EIS and LSV measurements. An amplitude of 10mV with 8 points/decade was used for the data collection, with a frequency range of 5 Mhz to 0.005 Hz for the full range EIS obtained before the LSV tests. This was fitted using ZView software.

*FIB:* Thermofisher Scios 2 DualBeam was used for FIB cutting of the LLZTO to reveal the interface. Using a homemade air-tight holder, air exposure was minimized. FIB cross-sectioning was done at 30 kV and 30 nA, with further polishing at 13 nA. The SEM image was obtained at 2 kV to minimize artifacts.

**Acknowledgment**

P.B. acknowledges the faculty startup support from Washington University in St. Louis. The materials characterization experiments were partially supported by IMSE (Institute of Materials Science and Engineering) at Washington University in St. Louis. The authors thank Dr. James Schilling and Dr. James Buckley for the help with the diamond saw.

**Author contributions**

P.B. designed and supervised the study and led the theoretical analysis. P.B., L.W., and R.G. synthesized the LLZTO powders and sintered the pellets. R.G. fabricated the cells, performed the electrochemical tests and microscopy characterization, and analyzed the data. J.G. contributed to the initial experiments. Y.L. performed the XRD characterization. P.B. and R.G. wrote and revised the manuscript.

**Additional information**

Supplementary information is available online.

**Competing interests**

The authors declare no competing financial interests.

**Data availability**

The data that support the findings of this study are available from the corresponding author upon reasonable request.

**Code availability**

Not applicable for this study.